\documentclass[12pt]{article}
\usepackage{amsmath,amsthm,latexsym,amssymb,amsfonts,epsfig}
\title{On the quantum theory of massless spin-3/2 field in Minkowski spacetime}
\author{\begin{tabular}{c}
\bigskip Weigang Qiu\footnote{Email: wgqiu@hutc.zj.cn}\\
Department of Physics, Huzhou Teachers College, Zhejiang, 313000,
China\\
\\
\bigskip Fei Sun\footnote{Email: Phaethonmonic@yahoo.com}\\
School of Mathematics, Peking University, Beijing, 100871, China\\
\\
\bigskip Hongbao Zhang\footnote{Email: hbzhang@pkuaa.edu.cn}\\
Department of Astronomy, Beijing Normal University, Beijing,
100875, China\\
Department of Physics, Beijing Normal University, Beijing, 100875,
China\\
CCAST(World Laboratory), P.O. Box 8730, Beijing, 100080,
China\end{tabular}}
\begin{document}
\maketitle
\begin{abstract}
From the modern viewpoint and by the geometric method, this paper
provides a concise foundation for the quantum theory of massless
spin-3/2 field in Minkowski spacetime, which includes both the
one-particle's quantum mechanics and the many-particle's quantum
field theory. The explicit result presented here is useful for the
investigation of spin-3/2 field in various circumstances such as
supergravity, twistor programme, Casimir effect, and quantum
inequality.
\end{abstract}
\section{Introduction}
As is well known, relativistic quantum theory originates from the
natural marriage of special relativity and quantum theory.
According to the modern viewpoint, the Hilbert space for one
particle quantum wave functions forms the unitary representation
of the Poincare group, which is the isometric transformation group
of the Minkowski spacetime. Especially, as realized on the
Minkowski spacetime, the quantum wave functions need to satisfy
the linear field equation of
motion\cite{Wigner,BW,Geroch,Weinberg1,Weinberg2}. Furthermore,
the quantum field operator, which satisfies the same linear field
equation, is defined on the Fock space associated with the Hilbert
space of one particle states.

Obviously, the spin-3/2 field occupies a special position in our
attempts to understand nature both relativistically and quantum
mechanically. It is the spin-3/2 field that turns out to be the
simplest nontrivial higher spin field, which but plays a
significant role in supergravity and twistor programme. By the
geometric method, this paper is mainly intended to revisit the
quantum theory of massless spin-3/2 field from the modern
viewpoint mentioned in the beginning. In the next section, we
develop the one-particle's quantum mechanics for the massless
spin-3/2 field. the many-particle's quantum field theory for
massless spin-3/2 field is presented in Section 3. We conclude
with some applications and extensions in Section 4.
\section{One-particle's Quantum Mechanics for Massless Spin-3/2
Field}
\subsection{The Hilbert Space of One Particle States }
 Start with the free massless spin-3/2 field equation on the
Minkowski spacetime\cite{Wald}
\begin{eqnarray}
\nabla^{A'A}\phi_{ABC}=0,\label{Maxwell1}
\end{eqnarray}
where $\phi_{ABC}$ is a totally symmetric spinor field, called
field strength. It is obvious that the solutions to the above
equation form a complex vector space. To define an inner product
on our complex vector space, we first introduce the
Rarita-Schwinger potential field
$\psi^{A'}{_{BC}}$\cite{Penrose1}\footnote{It seems that this
potential field is misnamed Dirac-Fierz potential field by R.
Penrose in \cite{Penrose1}.}, i.e.,
\begin{equation}
\phi_{ABC}=\nabla_{AA'}\psi^{A'}{_{BC}},
\end{equation}
\begin{equation}
\psi^{A'}{_{BC}}=\psi^{A'}{_{(BC)}},\label{symmetry}
\end{equation}
then the field equation (\ref{Maxwell1}) can be rewritten in terms
of $\psi^{A'}{_{BC}}$ as
\begin{equation}
\nabla^{B'B}\psi^{A'}{_{BC}}=0.\label{Dirac}
\end{equation}
Whence a conserved current reads\cite{Sen,RS}
\begin{equation}
j_c[\psi,\psi']=\sqrt{2}\sigma_c{^{C'C}}\bar{\psi}^B{_{A'C'}}\psi'^{A'}{_{BC}},\label{current1}
\end{equation}
then the inner product can be defined as
\begin{equation}
(\phi,\phi')=(\psi,\psi')=\int_\Sigma
j^a[\psi,\psi']\epsilon_{abcd}.\label{inner1}
\end{equation}
Note that the conservation of $j_a[\psi,\psi']$ implies that this
inner product is independent of choice of the Cauchy surface
$\Sigma$\footnote{This point also means the unitarity of the
evolution of the free field.}. Thus, for the later convenience, we
choose the surface of the constant Lorentz coordinate time $x^0$ as
$\Sigma$ once and for all. Moreover, Eqn.(\ref{inner1}) can be
written as
\begin{equation}
(\phi,\phi')=(\psi,\psi')=\int_\Sigma (\frac{\partial}{\partial
x^0})^aj_a[\psi,\psi']\tilde{\epsilon}_{bcd},\label{inner2}
\end{equation}
where $\tilde{\epsilon}_{bcd}=(\frac{\partial}{\partial
x^0})^a\epsilon_{abcd}$ is the induced spatial volume element on
$\Sigma$.

In addition, by Eqn.(\ref{Dirac}), Eqn.(\ref{current1}), and
Eqn.(\ref{inner2}), the Stokes theorem shows that the inner product
is invariant under gauge transformations\cite{Penrose2}
\begin{eqnarray}
\psi^{A'}{_{BC}}&\rightarrow& \psi^{A'}{_{BC}}+\nabla^{A'}{_B}\varphi_C,\nonumber\\
\psi'^{A'}{_{BC}}&\rightarrow&
\psi^{A'}{_{BC}}+\nabla^{A'}{_B}\varphi'_C,
\end{eqnarray}
with
\begin{eqnarray}
\nabla^{A'A}\varphi_A&=&0,\nonumber\\
\nabla^{A'A}\varphi'_A&=&0.
\end{eqnarray}
Thus the Hilbert space of one particle states for massless
spin-3/2 field can be constructed by\footnote{More properly, one
should complete $H$ such that it can be called a genuine Hilbert
space. Actually, for the physical taste, we are a little sloppy
here, ignoring the rigorous mathematics. Fortunately, there is a
general completion procedure for obtaining a mathematically
precise Hilbert space along with a collection of operators on
it\cite{Geroch}.}
\begin{equation}
H=H^+\bigoplus\bar{H}^-,
\end{equation}
where $H^+$($H^-$) is the complex vector space of
positive(negative) frequency solutions to the field equation
(\ref{Maxwell1}) with respect to the Lorentz coordinate time
$x^0$, and $\bar{H}^-$ is the complex conjugation space of $H^-$.
That is, $\bar{H}^-$ is the complex vector space of positive
frequency solutions to the field equation
\begin{equation}
\nabla^{AA'}\phi_{A'B'C'}=0.\label{Maxwell2}
\end{equation}

\subsection{Conserved
Observables from the Poincare Lie Algebra} It is well known that
the Poincare Lie algebra can be realized by the Killing vector
fields on the Minkowski spacetime as follows
\begin{eqnarray}
P_\mu{^a}&=&i(\frac{\partial}{\partial x^\mu})^a,\nonumber\\
M_{\mu\nu}{^a}&=&i[x_\mu(\frac{\partial}{\partial
x^\nu})^a-x_\nu(\frac{\partial}{\partial x^\mu})^a].
\end{eqnarray}
According to the fact that the covariant derivative commutes with
the Lie derivatives via Killing vector fields, the operators from
the Poincare Lie algebra, i.e.
\begin{eqnarray}
\hat{P}^\mu\phi_{ABC}&=&\pounds_{P^\mu}\phi_{ABC},\nonumber\\
\hat{P}^\mu\phi_{A'B'C'}&=&\pounds_{P^\mu}\phi_{A'B'C'},\nonumber\\
\hat{M}_{\mu\nu}\phi_{ABC}&=&\pounds_{M_{\mu\nu}}\phi_{ABC},\nonumber\\
\hat{M}_{\mu\nu}\phi_{A'B'C'}&=&\pounds_{M_{\mu\nu}}\phi_{A'B'C'},
\end{eqnarray}
are well defined on(in) the Hilbert space of one particle states.
Later, employing the Leibnitz rule, the conservation of
$j_a[\psi,\psi']$, and the Stokes theorem, we find that the above
operators are hermitian with respect to the inner product
(\ref{inner2}). In addtion, since the inner product (\ref{inner2})
is independent of the choice of $\Sigma$, the above operators are
also conserved observables. Moreover, taking into account
$[\pounds_u, \pounds_v]=\pounds_{[u,v]}$ with $u$ and $v$
arbitrary vector fields, we can obtain
\begin{equation}
[\hat{P}_\mu,\hat{P}_\nu]=0,
\end{equation}
\begin{equation}
[\hat{P}_\mu,\hat{M}_{\rho\sigma}]=2i\eta_{\mu[\rho}\hat{P}_{\sigma]},
\end{equation}
\begin{equation}
[\hat{M}_{\mu\nu},\hat{M}_{\rho\sigma}]=2i(\eta_{\mu[\rho}\hat{M}_{\sigma]\nu}-\eta_{\nu[\rho}\hat{M}_{\sigma]\mu}).
\end{equation}
Here, $\hat{P}^\mu$ is the four-momentum operator. By the field
equation (\ref{Maxwell1}) and (\ref{Maxwell2}), we have
\begin{equation}
\hat{P}_\mu\hat{P}^{\mu}=-\Box=0,
\end{equation}
which shows that the eigenvalue of the four-momentum operator is
null. Furthermore, $\{\hat{L}_1\equiv \hat{M}_{23},\hat{L}_2\equiv
\hat{M}_{31},\hat{L}_3\equiv \hat{M}_{12}\}$ are the total angular
momentum operators, and $\{\hat{K}_1\equiv
\hat{M}_{01},\hat{K}_2\equiv \hat{M}_{02},\hat{K}_3\equiv
\hat{M}_{03}\}$ describe the uniform motion of center of mass.

We next introduce the Pauli-Lubanski spin vector operator
\begin{equation}
\hat{S}_\mu=\frac{1}{2}\epsilon_{\mu\nu\rho\lambda}\hat{P}^\nu
\hat{M}^{\rho\lambda},
\end{equation}
then we have
\begin{eqnarray}
\hat{S}^\mu\phi_{ABC}&=&\frac{1}{2}\epsilon^{\mu\nu\rho\lambda}\hat{P}_\nu
\hat{M}_{\rho\lambda}\phi_{ABC}\nonumber\\
&=&-\frac{1}{2}\epsilon^{\mu\nu\rho\lambda}(\frac{\partial}{\partial
x^\nu})^a\nabla_a\{[x_\rho(\frac{\partial}{\partial
x^\lambda})^b-x_\lambda(\frac{\partial}{\partial x^\rho})^b]\nabla_b\phi_{ABC}+3\phi_{D(BC}U_{|\rho\lambda|A)}{^D}\}\nonumber\\
&=&-\frac{1}{2}\epsilon^{\mu\nu\rho\lambda}\{[\eta_{\nu\rho}(\frac{\partial}{\partial
x^\lambda})^b-\eta_{\nu\lambda}(\frac{\partial}{\partial
x^\rho})^b]\nabla_b\phi_{ABC}\nonumber\\
&&+[x_\rho(\frac{\partial}{\partial
x^\nu})^a(\frac{\partial}{\partial
x^\lambda})^b-x_\lambda(\frac{\partial}{\partial
x^\nu})^a(\frac{\partial}{\partial
x^\rho})^b]\nabla_a\nabla_b\phi_{ABC}+3\phi_{D(BC}U_{|\rho\lambda|A)}{^D}\}\nonumber\\
&=&-\frac{3}{2}\epsilon^{\mu\nu\rho\lambda}(\frac{\partial}{\partial
x^\nu})^a\nabla_a[\phi_{D(BC}U_{|\rho\lambda|A)}{^D}].
\end{eqnarray}
Here, we have used $\phi_{ABC}=\phi_{(ABC)}$ in the second step, and
$\nabla_a\nabla_b=\nabla_b\nabla_a$ in the final step. In addition,
$U_{\rho\lambda A}{^D}$ reads\cite{GZ}
\begin{equation}
U_{\rho\lambda
A}{^D}=\frac{1}{2}\nabla_{AA'}M_{\rho\lambda}{^{DA'}}=\frac{1}{2}\nabla_{AA'}[x_\rho(\frac{\partial}{\partial
x^\lambda})^{DA'}-x_\lambda(\frac{\partial}{\partial
x^\rho})^{DA'}]=\frac{1}{2}(\sigma_{\rho
AA'}\sigma_\lambda{^{DA'}}-\sigma_{\lambda
AA'}\sigma_\rho{^{DA'}}).
\end{equation}
Then employing Eqn.(\ref{antisymmetric}) in Appendix B, we have
\begin{equation}
U_{\rho\lambda A}{^D}=\frac{i}{2}\epsilon_{\rho\lambda
AC'}{^{DC'}},
\end{equation}
thus
\begin{eqnarray}
\hat{S}^\mu\phi_{ABC}&=&-\frac{3i}{4}\epsilon^{\rho\lambda\mu\nu}\epsilon_{\rho\lambda
(A|C'|}{^{DC'}}(\frac{\partial}{\partial
x^\nu})^a\nabla_a\phi_{|D|BC)}\nonumber\\
&=&\frac{3i}{2}[\sigma^\mu{_{(A|C'|}}\sigma^{\nu
DC'}-\sigma^\nu{_{(A|C'|}}\sigma^{\mu
DC'}](\frac{\partial}{\partial
x^\nu})^a\nabla_a\phi_{|D|BC)}\nonumber\\
&=&\frac{3i}{2}[\sigma^\mu{_{(A|C'|}}\nabla^{DC'}\phi_{|D|BC)}-\sigma^{\mu
DC'}\nabla_{(A|C'|}\phi_{|D|BC)}]\nonumber\\
&=&-\frac{3i}{2}\sigma^{\mu
DC'}\nabla_{DC'}\phi_{(ABC)}=-\frac{3}{2}\hat{P}^\mu\phi_{ABC},
\end{eqnarray}
where we have used
$\epsilon^{abcd}\epsilon_{abef}=-4\delta^{[c}{_e}\delta^{d]}{_f}$
in the second step, and the field equation (\ref{Maxwell1}) has
been used in the forth step. The above equation implies that those
states in $H^+$ are the states with the helicity $-\frac{3}{2}$.
Similarly, we can obtain that those states in $\bar{H}^-$ are the
states with the helicity $\frac{3}{2}$, i.e.,
\begin{equation}
\hat{S}^\mu\phi_{A'B'C'}=\frac{3}{2}\hat{P}^\mu\phi_{A'B'C'}.
\end{equation}
\subsection{The Plane Wave Expansion Basis in the Coulomb Gauge}
In this section, we shall employ the Rarita-Schwinger potential
field in the Coulomb gauge to provide the complete orthonormal
expansion basis for the Hilbert space $H$ in the momentum
representation. To proceed, first let
\begin{equation}
\psi_a{^B}=\sigma_a{^{A'C}}\psi_{A'C}{^B},
\end{equation}
then Eqn.(\ref{Dirac}) can be written as
\begin{equation}
\sigma^a{_{B'B}}\nabla_a\psi_b{^B}=0,\label{Rarita}
\end{equation}
where we have used Eqn.(\ref{symmetry}), i.e.,
\begin{equation}
\sigma^a{_{B'B}}\psi_a{^B}=0.\label{Schwinger}
\end{equation}
Note that Eqn.(\ref{Rarita}) and Eqn.(\ref{Schwinger}) are just
the Rarita-Schwinger equations for massless spin-3/2 field, which
is the reason why the potential field here is called the
Rarita-Schwinger potential field.

Next in the coulomb gauge, i.e.,
\begin{equation}
(\frac{\partial}{\partial x^0})^a\psi_a{^B}=0,\label{Coulomb}
\end{equation}
and after a straightforward calculation, we obtain the complete
 plane wave solutions to Eqn.(\ref{Rarita}) and
Eqn.(\ref{Schwinger}) in the momentum representation as
\begin{equation}
\psi_p{_a{^B}}(x)=\frac{1}{\sqrt{(2\pi)^3}}\frac{1}{\sqrt{2|p_0|}}\tilde{\psi}_\mu{^\Sigma}(p)(dx^\mu)_a(\varepsilon_\Sigma)^Be^{-ip_bx^b}.
\end{equation}
Here
\begin{equation}
\tilde{\psi}(1,0,0,1)=(0,1,i,0)\otimes\left(\begin{array}{l l}
1\\0\end{array}\right),
\end{equation}
and
\begin{eqnarray}
\tilde{\psi}_\mu{^\Sigma}(p=e^{-\lambda},e^{-\lambda}\sin\theta\cos\varphi,e^{-\lambda}\sin\theta\sin\varphi,e^{-\lambda}\cos\theta)
&=&\tilde{\psi}_\mu{^\Sigma}(-p)\nonumber\\
&=&(\Lambda^{-1})^\nu{_\mu}L^\Sigma{_\Gamma}\tilde{\psi}_\nu{^\Gamma}(1,0,0,1),
\end{eqnarray}
where
\begin{eqnarray}
\Lambda^\mu{_\nu}&=&\left(\begin{array}{cccc}
         1 & 0 & 0 & 0 \\
          0 & \cos\varphi & -\sin\varphi & 0 \\
          0 & \sin\varphi & \cos\varphi & 0 \\
          0 & 0 & 0 & 1
       \end{array}\right)
\left(\begin{array}{cccc}
         1 & 0 & 0 & 0 \\
          0 & \cos\theta & 0 & \sin\theta \\
          0 & 0 & 1 & 0\\
          0 & -\sin\theta & 0 & \cos\theta
       \end{array}\right)
\left(\begin{array}{cccc}
         \cosh\lambda & 0 & 0& -\sinh\lambda \\
          0 & 1 & 0 & 0 \\
          0 & 0 & 1 & 0 \\
          -\sinh\lambda & 0 & 0 & \cosh\lambda
       \end{array}\right),\nonumber\\
L^\Sigma{_\Gamma}&=&\left(\begin{array}{cc}
        e^{-i\frac{\varphi}{2}} & 0 \\
          0 & e^{i\frac{\varphi}{2}}
       \end{array}\right)
\left(\begin{array}{cc}
        \cos\frac{\theta}{2} & -\sin\frac{\theta}{2} \\
          \sin\frac{\theta}{2} & \cos\frac{\theta}{2}
       \end{array}\right)
\left(\begin{array}{cc}
        e^{-\frac{\lambda}{2}} & 0 \\
          0 & e^{\frac{\lambda}{2}}
       \end{array}\right).
\end{eqnarray}
Thus employing the inner product (\ref{inner2}) with the conserved
current
\begin{equation}
j_c[\psi,\psi']=-\sqrt{2}\sigma_{cC'C}\bar{\psi}_a{^{C'}}\psi'^{aC},\label{current2}
\end{equation}
it can be shown that
\begin{eqnarray}
\psi_p{_a{^B}}(x)\in H^+, &&p_0>0, \nonumber\\
\psi_p{_a{^{B'}}}(x)=\bar{\psi}_{-p}{_a{^{B'}}}(x)\in\bar{H}^-,
&&p_0>0
\end{eqnarray}
forms the complete orthonormal expansion basis for $H$.
\section{Many-particle's Quantum Field Theory for Massless Spin-3/2
Field}
\subsection{The Quantum Field Operator of Many Particles System}
Let $F_a(H)$ be the anti-symmetric Fock space associated with $H$.
the annihilation and creation operators are defined on $F_a(H)$ as
usual\cite{Geroch,Wald}. Then the quantum field operator is
constructed as
\begin{equation}
\hat{\psi}_a{^B}(x)=\varrho_{Ia}{^B}(x)a(\bar{\varrho}^I)+c^\dag(\tau_I)\bar{\tau}^I{_a{^B}}(x),
\end{equation}
where $\{\varrho_I\}$ and $\{\tau_I\}$ are the complete
orthonormal bases of $H^+$ and $\bar{H}^-$ respectively. It can be
shown that the quantum field operator constructed above is
independent of the choice of the complete orthonormal basis. In
particular, by the plane wave expansion basis, we have
\begin{equation}
\hat{\psi}_a{^B}(x)=\int
d^3\mathbf{p}[a(\mathbf{p})\psi_p{_a{^B}}(x)+c^\dag(\mathbf{p})\psi_{-p}{_a{^B}}(x)],p_0>0,
\end{equation}
where the annihilation and creation operators satisfy the
anti-commutation relations as follows
\begin{eqnarray}
\{a(\mathbf{p}),a(\mathbf{p'})\}&=&0,\nonumber\\
\{a(\mathbf{p}),a^\dag(\mathbf{p'})\}&=&\delta^3(\mathbf{p}-\mathbf{p'}),\nonumber\\
\{a^\dag(\mathbf{p}),a^\dag(\mathbf{p'})\}&=&0,\nonumber\\
\{c(\mathbf{p}),c(\mathbf{p'})\}&=&0,\nonumber\\
\{c(\mathbf{p}),c^\dag(\mathbf{p'})\}&=&\delta^3(\mathbf{p}-\mathbf{p'}),\nonumber\\
\{c^\dag(\mathbf{p}),c^\dag(\mathbf{p'})\}&=&0.
\end{eqnarray}
\subsection{Energy Momentum Tensor via the Belinfante's Construction}
In order to construct the energy momentum tensor for massless
spin-3/2 field, we here resort to the Rarita-Schwinger
Lagrangian\cite{RS}
\begin{equation}
\mathcal{L}=-i\sqrt{2}[\bar{\psi}^{aB'}\sigma^b{_{B'B}}\nabla_b\psi_a{^B}-\frac{1}{3}(\bar{\psi}^{aB'}\sigma_{aB'B}\nabla_b\psi^{bB}+\bar{\psi}^{aB'}\sigma_{bB'B}\nabla_a\psi^{bB})+\frac{2}{3}\bar{\psi}^{aB'}\sigma_{aB'B}\sigma^{bBC'}\sigma_{cC'C}\nabla_b\psi^{cC}].\label{Lagrangian}
\end{equation}
Since the Belinfante's energy momentum tensor is equivalent with the
metric energy momentum tensor, we here employ the Belinfante's
energy momentum tensor constructed by\cite{Zhang1}
\begin{equation}
T_\mathcal{B}^{ab}=T_\mathcal{C}^{(ab)}+\nabla_cN^{(ab)c},
\end{equation}
where the canonical energy momentum tensor
\begin{equation}
T_\mathcal{C}^{ab}=\frac{\partial\mathcal{L}}{\partial\nabla_a\psi_d{^D}}\nabla^b\psi_d{^D}-\mathcal{L}\eta^{ab},
\end{equation}
and
\begin{equation}
N^{abc}=\frac{\partial\mathcal{L}}{\partial\nabla_a\psi_d{^D}}[(\delta_d{^b}\psi^{cD}-\delta_d{^c}\psi^{bD})-\frac{1}{2}(\sigma^b{_{EE'}}\sigma^{cDE'}-\sigma^c{_{EE'}}\sigma^{bDE'})\psi_d{^E}].
\end{equation}
Then by the Rarita-Schwinger equation and Eqn.(\ref{Ghost}) in
Appendix B, the Belinfante's energy momentum tensor reads
\begin{equation}
T_\mathcal{B}^{ab}=-i\sqrt{2}[\frac{1}{2}(\bar{\psi}^{dD'}\sigma^{(b}{_{D'E}}\nabla^{a)}\psi_d{^E}-\nabla^{(a}\bar{\psi}^{|dD'|}\sigma^{b)}{_{D'E}}\psi_d{^E})+(\nabla_c\bar{\psi}^{(b|D'|}\sigma^{a)}{_{D'D}}\psi^{cD}-\bar{\psi}^{cD'}\sigma^{(a}{_{D'D}}\nabla_c\psi^{b)D})].
\end{equation}
Furthermore, according to \cite{Liang}, given a Killing vector field
$\xi$ in the Minkowski spacetime, we have
\begin{equation}
\int_\Sigma
T_\mathcal{B}^{ab}\xi_b\epsilon_{aefg}=\int_\Sigma(\frac{\partial\mathcal{L}}{\partial\nabla_a\psi_d{^D}}\pounds_{\xi}\psi_d{^D}-\xi^a\mathcal{L})\epsilon_{aefg}=\int_\Sigma
j^a[\psi, i\pounds_{\xi}\psi]\epsilon_{aefg},
\end{equation}
which is obviously gauge invariant. Moreover, it can be obtained
that
\begin{equation}
\int_\Sigma
:\hat{T}_\mathcal{B}^{ab}\xi_b:\epsilon_{aefg}=a^\dag(\rho_I)a(\bar{\rho}^J)\lceil
i\pounds_\xi\rfloor^I{_J}+c^\dag(\tau_I)c(\bar{\tau}^J)\lfloor
i\pounds_\xi\rceil^I{_J},
\end{equation}
where
\begin{eqnarray}
\lceil i\pounds_\xi\rfloor^I{_J}&=&(\rho_I, i\pounds_\xi\rho_J), \nonumber\\
\lfloor i\pounds_\xi\rceil^I{_J}&=&(\tau_I, i\pounds_\xi\tau_J).
\end{eqnarray}
Especially, we have
\begin{equation}
\int d^3\mathbf{x}:\hat{T}^{0\mu}:=\int
d^3\mathbf{p}p^\mu(a^\dag(\mathbf{p})a(\mathbf{p})+c^\dag(\mathbf{p})c(\mathbf{p})),
\end{equation}
which is our familiar result.
\section{Discussions}
The result obtained here provides a basis for us to investigate
the Casimir effect and quantum inequality for massless spin-3/2
field, which has be reported elsewhere\cite{LXZ,HLZ}. In addition,
we would like to stress that the framework and method presented
here are also applicable to other particles with arbitrary mass
and spin such as photon\cite{HQZ}.
\section*{Acknowledgements}
We are paticularly grateful to Prof. R. P. Geroch for his
interesting exchanges of ideas and unpublished lecture notes,
which directly stimulates our investigation of this project. In
addition, we also thank Prof. P. van Nieuwenhuizen and Prof. H. Yu
for private communications on the Lagrangian of massless spin-3/2
field. H. Zhang would like to thank the Center for Gravity and
Relativistic Astrophysics at Nanchang University for its
hospitality during his present visit. W. Qiu's work was supported
by NSFC(Grant 10547116), the Science Research Fund of Huzhou
Teachers College(No.KX21001) and the Science Research Fund of
Huzhou City(No.KY21022).  H. Zhang's work was supported in part by
NSFC(Grant 10205002), NSFC(Grant 10373003), and NSFC(Grant
10533010).
\section*{Appendix A: Notations and Conventions}
Our notations and conventions follow those of \cite{Geroch}. In
particular, the ordinary vector fields are related with the spinor
fields by the soldering form $\sigma_a{^{AA'}}$. For example, the
Minkowski metric
$\eta_{ab}=\sigma_a{^{AA'}}\sigma_b^{{BB'}}\epsilon_{AB}\epsilon_{A'B'}$,
the covariant derivative $\nabla_a=\sigma_a{^{AA'}}\nabla_{AA'}$,
and the volume element compatible with the metric
$\epsilon_{abcd}=\sigma_a{^{AA'}}\sigma_b{^{BB'}}\sigma_c{^{CC'}}\sigma_d{^{DD'}}\epsilon_{AA'BB'CC'DD'}$
with
$\epsilon_{AA'BB'CC'DD'}=i(\epsilon_{AB}\epsilon_{CD}\epsilon_{A'C'}\epsilon_{B'D'}-\epsilon_{AC}\epsilon_{BD}\epsilon_{A'B'}\epsilon_{C'D'})$.
In addition, the index is raised or lowered by
$\{\epsilon_{AB},\epsilon_{A'B'},\eta_{ab}\}$. The d'Alembertian
is defined as $\Box=\nabla_a\nabla^a$. Furthermore, the Lorentz
coordinate system is specially denoted by $\{x^\mu|\mu=0,1,2,3\}$,
and the spatial vectors are indicated by letters in boldface.
Finally, the dyad spinor basis is denoted by
$\{(\varepsilon_\Sigma)^A|\Sigma=1,2\}$, where
\begin{eqnarray}
\epsilon_{\Sigma\Omega}=\epsilon^{\Sigma\Omega}&=&\left(\begin{array}{cc}
                           0 & 1 \\
                           -1 & 0
                         \end{array}\right),\nonumber\\
\sigma^\mu_{\Sigma'\Sigma}&=&\frac{1}{\sqrt{2}}(I,\sigma),\nonumber\\
\sigma^{\mu\Sigma\Sigma'}&=&\frac{1}{\sqrt{2}}(I,-\sigma),\nonumber\\
\eta_{\mu\nu}=\eta^{\mu\nu}&=&\left(\begin{array}{cccc}
                            1 & 0 & 0 & 0 \\
                            0 & -1 & 0 & 0 \\
                            0 & 0 & -1 & 0 \\
                            0 & 0 & 0 & -1
                          \end{array}
 \right),\nonumber\\
\nabla_\mu&=&\partial_\mu,\nonumber\\
\epsilon_{0123}&=&1.
\end{eqnarray}
\section*{Appendix B: Some Useful Identities}
Start with the spinor formulation of the volume element
\begin{equation}
\epsilon_{AA'BB'CC'DD'}=i(\epsilon_{AB}\epsilon_{CD}\epsilon_{A'C'}\epsilon_{B'D'}-\epsilon_{AC}\epsilon_{BD}\epsilon_{A'B'}\epsilon_{C'D'}),
\end{equation}
we have
\begin{eqnarray}
\epsilon_{abCC'D}{^{C'}}&=&\sigma_a{^{AA'}}\sigma_b{^{BB'}}\epsilon_{AA'BB'CC'D}{^{C'}}\nonumber\\
&=&i(\sigma_{aBC'}\sigma_b{^{BC'}}\epsilon_{CD}-2\sigma_{aCB'}\sigma_{bD}{^{B'}})\nonumber\\
&=&i(\eta_{ab}\epsilon_{CD}-2\sigma_{aCB'}\sigma_{bD}{^{B'}}).
\end{eqnarray}
Whence a pair of identities can be obtained as
\begin{eqnarray}
\sigma_{aCB'}\sigma_{bD}{^{B'}}+\sigma_{bCB'}\sigma_{aD}{^{B'}}&=&\eta_{ab}\epsilon_{CD},\nonumber\\\label{symmetric}
\sigma_{aCB'}\sigma_{bD}{^{B'}}-\sigma_{bCB'}\sigma_{aD}{^{B'}}&=&i\epsilon_{abCC'D}{^{C'}}.\label{antisymmetric}
\end{eqnarray}
Furthermore, from Eqn.(\ref{symmetric}), we have\cite{Han}
\begin{equation}
\sigma_{aAA'}\sigma^{cA'B}\sigma_{bBB'}+\sigma_{bAA'}\sigma^{cA'B}\sigma_{aBB'}=\delta_a{^c}\sigma_{bAB'}+\delta_b{^c}\sigma_{aAB'}-\eta_{ab}\sigma^c{_{AB'}}.\label{Ghost}
\end{equation}

\end{document}